\documentclass[aps,prl,twocolumn,groupedaddress,showpacs,floatfix]{revtex4}
\usepackage{graphicx}
\usepackage{amsmath}

\begin{document}

\title{Polarons in the radio-frequency\\ spectrum of a quasi-two-dimensional Fermi gas}

\author{Y. Zhang$^{1,2}$, W. Ong$^{1,2}$, I. Arakelyan$^{1,2}$, and J. E. Thomas$^{1}$}
\affiliation{$^{1}$Department of  Physics, North Carolina State University, Raleigh, NC 27695, USA}
\affiliation{$^{2}$Department of Physics, Duke University, Durham, NC 27708, USA}

\pacs{03.75.Ss}

\date{\today}

\begin{abstract}
We measure radio-frequency spectra for a  two-component mixture of a $^6$Li atomic Fermi gas  in the  quasi-two-dimensional regime.  Near the Feshbach resonance,  where the transverse Fermi energy is large compared to the confinement-induced dimer binding energies for the initial and final states, we find that the observed resonances  do not correspond to transitions between confinement-induced dimers. The spectrum  appears to be well-described by transitions between noninteracting polaron states in two dimensions.
\end{abstract}

\maketitle

Quantum degenerate atomic Fermi gases, with magnetically controlled interactions, are ideally suited for exploring pairing interactions in reduced dimensions~\cite{Randeria2D,Esslinger1DFermi,TormaTwoDim,Giorgini2DFermiCrossover,KohlRF2DSIFermi,Sommer2D}. Near a broad collisional (Feshbach) resonance, a two-component gas in three dimensions can be continuously tuned from a Bardeen-Cooper-Schrieffer (BCS) superfluid, which exhibits weakly bound Cooper pairs, to a resonant strongly interacting  superfluid and finally to a Bose-Einstein condensate (BEC) of molecular dimers. Since 2002,  degenerate strongly interacting Fermi gases have been studied in three-dimensional geometries, providing a paradigm for strongly interacting systems in nature, from high temperature superconductors to nuclear matter~\cite{OHaraScience,RMP2008,ZwergerReview,ZwierleinFermiReview}. In contrast to free space, where bound dimers exist only in the BEC regime,  two-dimensional (2D) confinement also stabilizes  bound dimers in the BCS regime~\cite{Randeria2D,PetrovShlyapnikov2D}.
The interplay between confinement-induced pairing and many-body physics in 2D confined mesoscopic systems of several hundred atoms has not been previously explored and offers new challenges for predictions~\cite{Randeria2D,TormaTwoDim,Giorgini2DFermiCrossover}.

We study  a  quasi-two dimensional Fermi gas of $^6$Li in the  many-body regime. Radio-frequency (rf) spectra are obtained for a 50-50 mixture of the two lowest hyperfine states denoted 1 and 2, by driving transitions to an initially empty hyperfine state 3 and measuring the depletion of state 2.  At 720 G, well-below the Feshbach resonance,  the molecular 12 dimer binding energy is  larger than the local Fermi energy, and the observed spectra exhibit the expected threshold form, arising from dissociation of 12 dimers into 13 scattering (``free") states. However, near the Feshbach resonance at 834 G the predicted spectra for transitions between dimer states are in marked disagreement with the measurements, where the transverse Fermi energy is larger than the 2D dimer binding energy. Here, we find that the resonance locations appear to be described by transitions between noninteracting polaron states, which describe an impurity atom in state 2 or in state 3, immersed in a bath of atoms in state 1. In this regime, polarons are expected to be energetically more favorable than the corresponding dimers~\cite{MartinPolaron,Pethick2DPolarons,Parish2DPolarons}.

Prior experiments have been performed with  a dilute gas, where the local transverse Fermi energy $\mu$ is small compared to the binding energy $E_b$ of a single dimer. In this case, the dimers are small compared to  the interparticle spacing and one expects the measured rf spectra to be consistent with predictions based on molecular binding in the BEC regime and confinement-induced dimers in the BCS regime, as observed previously by measuring the threshold for free-to-bound transitions~\cite{KohlRF2DSIFermi} and  bound-to-free transitions~\cite{Sommer2D}.

Our experiments are performed instead in the many-body limit, where the the transverse chemical potential $\mu>E_b$ over most of the trap,  but $\mu\simeq 1.5\,h\nu_z$ is small enough to be in the quasi-two-dimensional regime~\cite{TormaTwoDim}. For a simple BCS approximation in two dimensions, the trap-averaged rf transition rate to excite an atom from one populated state in a 50-50 mixture to an unpopulated noninteracting final state is $\propto \int_0^\infty d^2\mathbf{x}_\perp |\Delta|^2 \theta [\hbar\omega+\mu-\sqrt{\mu^2+|\Delta|^2}]/\omega^2$, where $\omega$ is the radio frequency relative to the bare atomic transition frequency and $\Delta$ is the pairing gap for the initial mixture. Here, we assume for simplicity that the temperature $T<<\mu$   over most of the trap, so that the number of excitations is negligible. For $\mu>>E_b$, predictions for a 2D gas~\cite{Randeria2D} give $\Delta=\sqrt{2\mu\,E_b}$ in a mean field (BCS) approximation, yielding a threshold for the bound (Cooper pair)-to-free spectrum at $\hbar\omega\simeq \Delta^2/(2\mu)=E_b$, which is just the dimer binding energy, as noted previously~\cite{Sommer2D}. Hence, the predicted  spectra are identical with the dilute gas limit, in contrast to our data taken in the same regime.

Table~\ref{table:1} lists the experimental parameters for several different magnetic fields and trap depths.
The CO$_2$ laser standing-wave trapping potential is characterized by using parametric resonance in the weakly interacting regime near 300 G  to determine the oscillation frequencies of the atoms in the transverse  directions ($\nu_x$,$\nu_y$) and in the tightly confined axial direction ($\nu_z$). The axial trapping potential is taken to be $U_{axial}=U_0\,\sin^2(2\pi z/\lambda)$, where the trap depth is readily determined from $\nu_z$ in the harmonic approximation, $U_0=m(\nu_z\lambda)^2/2$.  We find  $\nu_{\perp}\equiv\sqrt{\nu_x\nu_y}=\nu_z/25$.   The number of atoms per site is estimated using the lattice spacing of $5.3\,\mu$m and the number of atoms in the central part (along $z$) of the cloud, as measured by absorption imaging.
 \begin{table}
\begin{tabular}{|c|c|c|c|c|c|c|c|}
  \hline
  $B(G)$ & $U_0$($\mu$K)& $\nu_z$(kHz)& $N_{site}$ & $E_{F\perp}$($\mu$K)  & $E_b^{12}$ & $E_b^{13}$ &  $\epsilon_{bb}$ \\
  \hline
  719 & 27.5 & 26.0 & 1298 & 1.87 & 145 & 2.9 & 0.27\\
  \hline
  809 & 23.5 & 24.0 & 1951 & 2.12 & 12.6 & 0.88 & 0.53\\
  \hline
  810 & 294 & 85 & 1134 & 5.73 & 33.1 & 6.69 &0.79\\
  \hline
  832 & 24.4 & 24.5 &1620 &1.97 & 7.25 & 0.81 & 0.66 \\
  \hline
  832 & 274 & 82.0 & 1500&6.36 & 23.9 & 5.79 & 0.83 \\
  \hline
  832 & 742 & 135 &1800 &11.47 & 39.1 & 12.45 & 0.89 \\
  \hline
  831 & 1304 & 179 &1250 &12.65 & 51.9 & 18.9 & 0.91 \\
  \hline
  842 & 24.4  & 24.5   &  1420    &  1.85    &  5.91    & 0.78    & 0.70\\
  \hline
  842 &  277   & 82.5   &  1617    &  6.64    &  21.4    & 5.66    & 0.85\\
  \hline
\end{tabular}
\caption{Parameters for the radiofrequency spectra: Trap depth $U_0$; Axial frequency $\nu_z$; Total number of atoms per pancake trap $N_{site}$; Transverse Fermi energy $E_{F\perp}=h \nu_\perp\sqrt{N_{site}}$; $E_b^{12}$ and $E_b^{13}$ are the dimer binding energies in kHz, for $\nu_{\perp}/\nu_z=1/25$.  The dimer binding energies are obtained using the Green's function method described in the supplementary material~\cite{Supplementary}. $\epsilon_{bb}$ is the corresponding bound dimer to bound dimer transition fraction (area under the bound to bound spectrum) determined from the binding energies obtained with the scattering lengths measured in Ref.~\cite{BartensteinFeshbach}.}
\label{table:1}
\end{table}

To compare the measured spectra to predictions, we begin by determining the 2D dimer  binding energies, $E_b^{ij}\equiv\epsilon_b^{ij}\,h\nu_z$ for atoms in states $i$ and $j$, as described in the supplementary material~\cite{Supplementary} and given in Table~\ref{table:1}. The calculation includes the finite transverse confinement of the trapping potential, where $\nu_\perp/\nu_z=1/25$,  which significantly increases the dimer binding energy, especially for weakly bound dimers.

The contribution to the spectrum from a dimer-to-dimer transition is determined by computing the corresponding fraction $\epsilon_{bb}$. In the weak binding approximation, including only the axial ground state part of the dimer wavefunction,  we obtain,
\begin{equation}
\int d\nu\,I_{bb}(\nu)\equiv\epsilon_{bb}(q)= \frac{q^2}{4 \sinh^2(q/2)},
\label{eq:boundtoboundfraction}
\end{equation}
where $q\equiv \ln(\epsilon_b^{13}/\epsilon_b^{12})$ for a $2\rightarrow 3$ transition in a 12 mixture.
We plot $\epsilon_{bb}$ as a function of magnetic field in the supplementary material~\cite{Supplementary}. There we show that similar results for $\epsilon_{bb}$ are obtained including the first 50 even axial states, which determines $\epsilon_{bb}$ for both weak and tight binding of the $1-2$ or $1-3$ dimers.

\begin{figure}[htb]
\includegraphics[width=3.25 in]{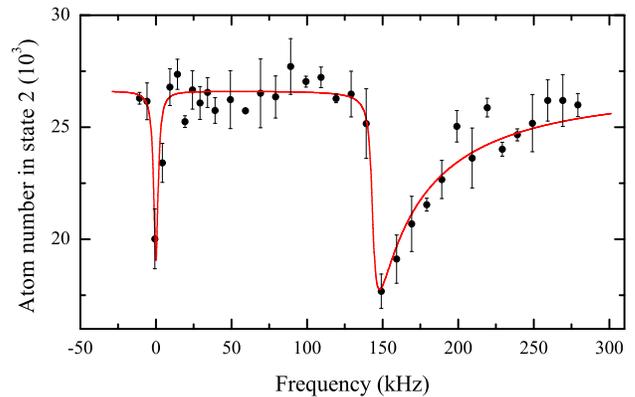}
\caption{RF spectra in a 1-2 mixture for a $2\rightarrow 3$ transitions in a quasi-two-dimensional $^6$Li Fermi gas near 720 G. The left resonance occurs at the bare atomic transition frequency and the threshold resonance spectrum on the right is in very good agreement with predictions for molecular dimers. \label{fig:720}}
\end{figure}

Fig.~\ref{fig:720} shows the measured spectra at 720 G, well below the Feshbach resonance, where the molecular dimer binding energy is larger than the local Fermi energy. At this field and trap depth, where the bound-to-bound transition fraction $\epsilon_{bb}=0.27$, we expect bound-to-free transitions to dominate the resonance spectrum~\cite{Supplementary}.
The resonance near 150 kHz is well fit by a threshold function for a quasi-two-dimensional gas using the calculated 12 and 13 dimer binding energies~\cite{Supplementary},
\begin{equation}
I_{bf}(\nu)=\frac{\epsilon_b^{12}\nu_z}{\nu^2}\,
\frac{q^2\,\theta(\nu-\epsilon_b^{12}\nu_z)}{\left[q-\ln\left(\frac{\nu}{\epsilon_b^{12}\nu_z}-1\right)\right]^2+\pi^2},
\label{eq:boundfree}
\end{equation}
where $\nu$ is the rf frequency in Hz, relative to the bare atomic transition frequency, and $E_b^{12}=\epsilon_b^{12}h\nu_z$ is the 12 dimer binding energy in Hz. Note that $\int d\nu\,I_{bf}(\nu)=1-\epsilon_{bb}(q)$, as it should.

\begin{figure}[htb]
\includegraphics[width=3.25 in]{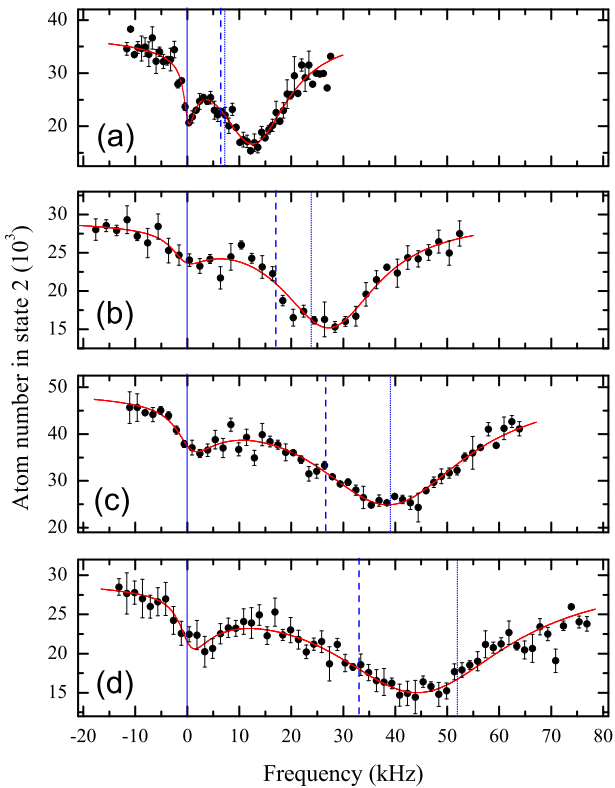}
\caption{RF spectra for a 1-2 mixture ($12\rightarrow13$ transition) near the Feshbach resonance at 834 G, versus trap depth $U_0$. The atom number remaining in state 2 is shown versus rf frequency: a) $U_0=21\,\mu$K, $\nu_z=24.5$ kHz; b) $U_0=280\,\mu$K, $\nu_z=82.5$ kHz;
c) $U_0=742\,\mu$K, $\nu_z=135$ kHz; d) $U_0=1304\,\mu$K, $\nu_z=179$ kHz.  Vertical lines show the measured bare atomic resonance position (solid) and the predicted frequencies for confinement-induced dimers: bound-to-bound transition resonance $h\nu=E_b^{12}-E_b^{13}$ (dashed) and the threshold for the bound-to-free transition  $h\nu =E_b^{12}$ (dotted).  According to Table~\ref{table:1}, the  bound-to-free transition (dotted line) should make a negligible contribution to the spectrum.\label{fig:834}}
\end{figure}

\begin{figure}[htb]
\includegraphics[width=3.25 in]{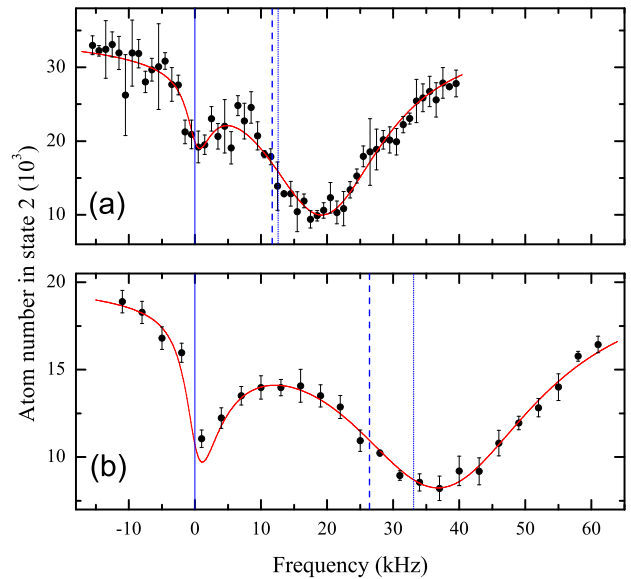}
\caption{RF spectra for $2\rightarrow 3$ transitions in a quasi-two-dimensional $^6$Li Fermi gas at 810 G: a) $U_0=23.5\,\mu$K, $\nu_z=24.0$ kHz; b) $U_0=294\,\mu$K, $\nu_z=85.0$ kHz.  Vertical lines show the measured bare atomic resonance position (solid) and the predicted frequencies for confinement-induced dimers: bound-to-bound transition resonance $h\nu=E_{b12}-E_{b13}$ (dashed) and the threshold for the bound-to-free transition  $h\nu =E_{b12}$ (dotted).\label{fig:811}}
\end{figure}

\begin{figure}[htb]
\includegraphics[width=3.25 in]{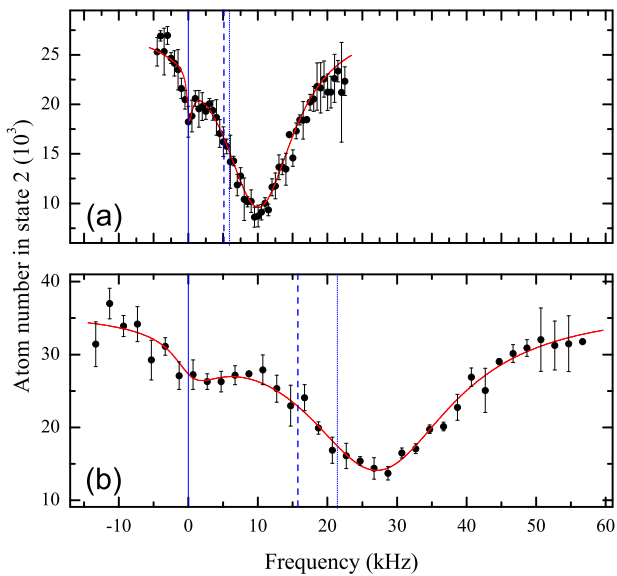}
\caption{RF spectra for $2\rightarrow 3$ transitions in a quasi-two-dimensional $^6$Li Fermi gas near 842 G: a) $U_0=24.4\,\mu$K, $\nu_z=25.0$ kHz; b) $U_0=277\,\mu$K, $\nu_z=82.5$ kHz.  Vertical lines show the measured bare atomic resonance position (solid) and the predicted frequencies for confinement-induced dimers: bound-to-bound transition resonance $h\nu=E_{b12}-E_{b13}$ (dashed) and the threshold for the bound-to-free transition  $h\nu =E_{b12}$ (dotted).\label{fig:842}}
\end{figure}

 Next, we examine spectra for the 12 mixture near the Feshbach resonance, at 832 G, Fig.~\ref{fig:834}, at 810 G, Fig.~\ref{fig:811}, and at 842 G, Fig.~\ref{fig:842}.
 For several of the spectra, the bare-atom resonance peak  on the left side exhibits a fast rise and a tail toward higher frequency, which we assume arises from a density-dependent mean field shift~\cite{LangmackClockshift}. At 832 G, the sharp threshold remains at nearly the same frequency (which we have set to 0 in Fig.~\ref{fig:834}), even when the trap depth is increased to maximum, where the ideal gas Fermi energy at the trap center is $E_{F\perp}=12.7\,\mu$K, i.e., $\simeq 260$~kHz. We assume that the threshold location is determined by the lowest density region at the cloud edges, where the mean field shift is negligible. The observed bare-atom resonance frequency is in agreement with that obtained at high temperature as well as for $2\rightarrow 3$ transitions with all the atoms initially in the $2$ state.

  Near 834 G, for the conditions of our experiments,  the bound-to-bound transition fractions in Table~\ref{table:1} are close to unity~\cite{Supplementary} so that bound-to-bound transitions should dominate the spectrum above the bare atomic resonance frequency.
 To determine  the frequency shift $\Delta\nu$ between the bare atom resonance and the second resonance peak to the right of it, we  fit a threshold line shape to the atomic peak, which is convolved with a narrow Lorenztian. In this case, the peak position is shifted to the right of the bare atom threshold location, which is determined from the fit.  A second Lorentzian, with a larger width, is fit to the shifted resonance peak. The results for $\Delta\nu$ are given in Table~\ref{table:2} and do not agree with the predictions for dimer-to-dimer transitions,  $h\Delta\nu_{dimer}=E_b^{12}-E_b^{13}$, shown as dashed lines in Figs.~\ref{fig:834}-\ref{fig:842}.

 We consider the possibility that polarons and not dimers determine the primary spectral features, so that the difference between the  initial and final state polaron energies determines the observed frequency shifts. We assume that the coherent part of the spectrum is approximately given by $Z\,\delta [\hbar\omega-E_p(3,1)+E_p(2,1)]$, where $E_p(i,1)$ is the polaron energy for an impurity atom in state $i=2,3$ immersed in a bath of state $1$. Since the momentum does not change in the rf transition $Z\simeq Z_2\,Z_3$ is determined by the overlap of the initial and final polaron momentum space wavefunctions~\cite{Supplementary}. For experiments at 832 G, we find that the individual polaron quasiparticle weights are $Z_2>0.8$ and $Z_3>0.9$~\cite{Supplementary}, assuming that the local density does not change in the transition. Therefore, we expect strong overlap between the initial and final polaron states, so that  transitions between polaron states should make an important contribution to the spectrum.

 We determine the 2D polaron energies for an isolated impurity atom in state 3 or state 2 in a bath of atoms in state 1, using the method described in the supplementary material~\cite{Supplementary}. The method is based on that described for a 3D gas in the supplementary material of Schirotzek at al.~\cite{MartinPolaron}, which utilizes the zero momentum polaron wavefunction proposed by Chevy~\cite{ChevyPolaron}. We extend that method to the 2D problem, by renormalizing the interaction strength as in Refs.~\cite{Randeria2D,Pethick2DPolarons,Parish2DPolarons}. Using the calculated dimer binding energies to determine the polaron energies, we find that for an atom in state 3 at 832 G, the energy for an isolated polaron  is attractive and more than half of the Fermi energy, much larger than the corresponding dimer binding energy. For an impurity in state 2, for which the 12 scattering length diverges, the polaron energy is attractive and somewhat larger than the Fermi energy. In this case, the polaron is localized to approximately the interparticle spacing, but is not small compared to the interparticle spacing, as it is at 720 G.

 In calculating the polaron frequency shift, we assume the peak position is determined by a local Fermi energy  $E_F=\lambda_1\, E_{F\perp}$,  where $E_{F\perp}$ is the ideal gas global Fermi energy given in Table~\ref{table:1}.  Fixing $\lambda_1=0.67$, we obtain the polaron frequency shifts $h\Delta\nu_{polaron}\equiv E_p(3,1)-E_p(2,1)$, which  are compared to the measurements in Table~\ref{table:2}. For the data just below and just above the resonance, at 810 and 842 G, Table~\ref{table:2}, we find excellent agreement.  Further, at 832 G, the calculated polaron frequency shifts agree very well with the measured frequency shifts at all four trap depths, Table~\ref{table:2}, demonstrating nontrivial scaling with the trap depth and axial frequency.

\begin{table}
\begin{tabular}{|c|c|c|c|}
  \hline
  $B(G)$ & $\nu_z$(kHz) &  $\Delta\nu_{meas}$(kHz) & $\Delta\nu_{polaron}$(kHz)\\
  \hline
   809 & 24.0 & 18.7 & 18.3\\
 \hline
 810 & 85 & 37.1 & 37.0\\
 \hline
 842 & 24.5 & 10.1 & 9.7\\
 \hline
 842 & 82.5 & 27.2 & 26.7 \\
 \hline
 832 & 24.5 &  12.3 & 11.6 \\
 \hline
 832 & 82.0 &   28.3 & 29.1  \\
 \hline
 832 &135 &  38.8 & 42.8 \\
 \hline
 831 & 179 &   44.5 & 48.3 \\
 \hline
 \end{tabular}
\caption{Frequency shift $\Delta\nu$ between the bare atom peak and the second resonance peak for a 12 mixture near the Feshbach resonance at different trap depths. The corresponding axial trap frequency is $\nu_z$.   The measured values of $\Delta\nu$ are compared to the values calculated assuming a transition from a polaron in state 2 to a polaron in state 3, in a bath of atoms in state 1. For determination of the polaron frequency shift, the dimer binding energies are taken from Table~\ref{table:1} and the local Fermi energy is taken to be $E_F=0.67\,E_{F\perp}$, where $E_{F\perp}$ is the ideal gas global Fermi energy given in Table~\ref{table:1}.\label{table:2}}
\end{table}

  We have also obtained spectra for a 50-50  mixture of states 1 and 3 near the Feshbach resonance at 690 G, where we drive either the $13\rightarrow12$ transition or the $13\rightarrow23$ transition. In both cases, the binding energy of the final state 12 or 23 dimers is large compared to the local Fermi energy and suppresses the bound-to-bound transition probability. For the high density regime of the experiments at the lowest temperatures, the bound-to-free part of the spectrum is not separated from the bare atom transition peak, and probably contain a dimer to free contribution, while at higher temperatures, there appears to be a polaron peak. These spectra are consistent with the predictions for polaronic transitions. However, more theoretical work is needed to explain the detailed shapes of all of the measured spectra, which may be improved in future experiments by local measurements~\cite{Sommer2D}.

 In summary,  we have observed rf spectra in a dense quasi-2D Fermi gas regime near a Feshbach resonance,  where the dimer binding energies for both the initial and final states are smaller than the transverse Fermi energy. The spectra appear to be explained by transitions between noninteracting polaronic states, despite the 50-50 mixture employed in the experiments.  These results support the conjecture~\cite{MartinPolaron} that at low temperature, a strongly interacting Fermi gas in a balanced mixture of two spin states may be described as a gas of polaron pairs.

 This research is supported by the Physics divisions of the Army Research Office, the Air Force Office Office of Sponsored Research, and the National Science Foundation, and the Division of Materials Science and Engineering,  the Office of Basic Energy Sciences, Office of Science, U.S. Department of Energy. The authors are indebted to Thomas Sch\"{a}fer and Lubos Mitas for stimulating discussions.


\end{document}


\title{Supplementary Material: ``Polarons in the radio-frequency\\ spectrum of a quasi-two-dimensional Fermi gas"}

\author{Y. Zhang$^{1,2}$, W. Ong$^{1,2}$, I. Arakelyan$^{1,2}$, and J. E. Thomas$^{1}$}
\affiliation{$^{1}$Department of  Physics, North Carolina State University, Raleigh, NC 27695, USA}
\affiliation{$^{2}$Department of Physics, Duke University, Durham, NC 27708, USA}

\pacs{03.75.Ss}

\date{\today}

\begin{abstract}
In this supplementary material, we discuss the calculation of the radio-frequency spectra arising from confinement-induced dimers and polarons in a quasi-two-dimensional Fermi gas. We determine the dimer binding energies, including both the tight axial confinement and the nonzero transverse confinement. We provide the probabilities for dimer-to-dimer  transitions and the shape of the dimer-to-scattering state spectrum. We also find the energy and quasi-particle weights for polarons in the two-dimensional gas and the corresponding spectra for  polaron to polaron transitions.
\end{abstract}

\maketitle

We begin by reviewing briefly in \S~\ref{sec:dimer} the radio-frequency spectrum arising from confinement-induced pairs, including final state interactions, but ignoring many-body effects, using the method employed for the three-dimensional case by Chin and Julienne~\cite{Chin-JulienneRF}. We consider mixtures of the three lowest hyperfine states of $^6$Li, denoted $|1\rangle$, $|2\rangle$, $|3\rangle$. For the conditions of our experiments in a 12 mixture at 720 G, the observed $2\rightarrow 3$ threshold spectrum is well described by a 12-dimer-to-13-scattering-state transition. In contrast, at 834 G, the predicted dimer spectrum is in marked disagreement with the data. In particular, we find that the difference between the ground and excited state dimer energies is too small. In \S~\ref{sec:polaron} we determine the energies for noninteracting confinement-induced polarons. We find that the locations of the observed resonances  for a 12 mixture near 834 G  are well modeled  by the predicted energy difference between isolated state 2 polarons and state 3 polarons, in a bath of atoms in state 1.

\section{Confinement-induced dimers}
\label{sec:dimer}
A simple golden rule calculation gives the radio-frequency-induced transition rate out of the initial state to all possible final states $R_i(\omega_{rf})=\sum_FR_{f\leftarrow i}$, where $R_{f\leftarrow i}=(2\pi/\hbar)|\tilde{H}'_{fi}|^2\,\delta(E_f-E_i-\hbar\omega_{rf})$, with $\tilde{H}'_{fi}=\hbar\Omega_{fi}\langle F|I\rangle/2$. Here, $\Omega_{fi}$ is the Rabi frequency for changing the hyperfine state of a single atom from the chosen populated state ($i$) to the initially unpopulated state ($f$) and $\langle F|I\rangle$ is the overlap between the initial and final  wave-functions for the relative motion of the atom-pair. Since the center of mass energy does not change in the rf transition,  $E_f-E_i$ is the total change in the atomic hyperfine energy ($\equiv\hbar\omega_{fi}$) plus the change in the  energy of the relative motion of the pair $E_F-E_I $.  Since $\sum_{F}|\langle F|I\rangle|^2=1$, $\int d\omega_{rf}\,R_{i}(\omega_{rf})=(\pi/2)\Omega_{fi}^2$. We define a normalized spectrum $I(\omega)$  where $R_{i}(\omega_{rf})=(\pi/2)\Omega_{fi}^2\,I(\omega)$ and $\omega_{rf}=\omega_{fi}+\omega$, with $\omega$ the frequency relative to the (unshifted) free-atom hyperfine transition frequency. Then, $I(\omega)=\sum_F|\langle F|I\rangle|^2\,\hbar\delta(E_F-E_I-\hbar\omega)$ and $\int d\omega I(\omega)=1$.

To determine the spatial wavefunctions and the pair binding energies, we note that the range of the two-body interaction is small compared to the interparticle spacing as well as to the harmonic oscillator confinement scale $l_z\equiv\sqrt{\hbar/(m\omega_z)}$. In this case, interactions between atoms in two different spin states are well described by the s-wave pseudopotential in three dimensions~\cite{ZwergerReview}, $V(r)=(4\pi\hbar^2a/m)\,\delta(r)\partial_r(r...)$, where $r$ is the distance between the atoms, $m$ is the mass of a single atom and $a$ is the magnetically tunable s-wave scattering length. The spatial wavefunctions are readily written in terms of the Green's function $G_E(\mathbf{r})$ for the relative motion of the two atoms in the confining potential, which we take to be harmonic, with ground state energy $E_0$. The two-atom scattering states with energy $E=E_0+E_s$, where $E_s\geq 0$, take the form $\psi_s(\mathbf{r})=\psi_{E_s}^{(0)}(\mathbf{r})-a\,G_{E_s}(\mathbf{r})\,u'_s(0)$, where $\psi_{E_s}^{(0)}$ is the input state and $\psi =u/r$ with $u$ regular at $r=0$. For the bound states, where there is no input, we have $\psi_b(\mathbf{r})=-a\,G_{E_b}(\mathbf{r})\,u'_b(0)$, where $E=E_0-E_b$ with $E_b > 0$. Using $\partial_r[r\psi_b(r)]_{r\rightarrow 0}=u'_b(0)$ yields the equation for the binding energy~\cite{ZwergerReview}, $1=-a\,\partial_r[rG_{E_b}(\mathbf{r})]_{r\rightarrow 0}$, where the right side projects out the regular part of $G$ at $r=0$.   For a three dimensional harmonic trap, the Green's function is
\begin{equation}
G_\epsilon(\mathbf{r})=\frac{i}{l_z\sqrt{4\pi}}\int_0^\infty \hspace*{-0.125in} d\eta\,e^{i\epsilon\eta}\,\prod_j\,e^{i\cot(\beta_j\eta)\left(\frac{x_j}{2l_j}\right)^2}f_j(\eta),
\label{eq:boundstate}
\end{equation}
where $E=E_0+\epsilon\,\hbar\omega_z$, with $\epsilon=-\epsilon_b$ for bound states and $\epsilon>0$ for scattering states. Here, $f_j(\eta)= \sqrt{2\beta_j/(1-e^{-2i\beta_j\eta})}$, with $\beta_z\equiv 1$, $\beta_{x,y}=\omega_{x,y}/\omega_z$ and $l_j=l_z/\sqrt{\beta_j}$.
After subtracting the irregular part of $G$ and changing to real variables, we determine the dimer binding energies $E_b=\epsilon_b\hbar\omega_z>0$ from
\begin{equation}
\hspace*{-0.02in}\frac{l_z}{a}=\int_0^\infty\hspace*{-0.125in}\frac{du}{\sqrt{4\pi u^3}}\left\{1-\prod_j\left(\frac{2\beta_ju}{1-e^{-2\beta_ju}}\right)^{1/2}\,e^{-\epsilon_bu}\right\}.
\label{eq:bindingenergy}
\end{equation}
 The dimer binding energy is significantly increased for nonzero transverse confinement. At resonance, where $l_z/a\rightarrow 0$, for $\nu_\perp/\nu_z=0$ we obtain $E_b=0.245\,h\nu_z$, while for $\nu_\perp/\nu_z=1/25$ we obtain $E_b=0.290\,h\nu_z$. At 842 G in the shallowest trap, the binding energy of the 13 dimer is increased from 0.15 kHz without transverse confinement to 0.78 kHz with transverse confinement. We compute the pair binding energy $E_b\equiv\epsilon_b\,\hbar\omega_z$  as a function of magnetic field using the s-wave scattering lengths $a$ measured in Ref.~\cite{BartensteinFeshbach}.

The scattering state is determined using $u'_s(0)={u'}_s^{(0)}(0)/\{1+a\,\partial_r[rG_{E_s}(\mathbf{r})]_{r\rightarrow 0}\}$. Using the relation (one-to-one correspondence) between the scattering length and the  bound state energy, we have $u'_s(0)={u'}_s^{(0)}(0)/\{a\,[G_{E_s}(\mathbf{r})-G_{E_b}(\mathbf{r})]_{r\rightarrow 0}\}$, where $E_b$ is the binding energy corresponding to the scattering length $a$ and we have used $\partial_r \{r[G_{E_s}(\mathbf{r})-G_{E_b}(\mathbf{r})]_{r\rightarrow 0}\}=[G_{E_s}(\mathbf{r})-G_{E_b}(\mathbf{r})]_{r\rightarrow 0}$, which is regular at $r=0$. Then, the scattering state takes the form
\begin{equation}
\psi_s(\mathbf{r})=\psi_{E_s}^{(0)}(\mathbf{r})-\frac{G_{E_s}(\mathbf{r})\,{u'}_s^{(0)}(0)}{[G_{E_s}(\mathbf{r})-G_{E_b}(\mathbf{r})]_{r\rightarrow 0}},
\label{eq:scattering}
\end{equation}
where ${u'}_s^{(0)}(0)=\psi_{E_s}^{(0)}(0)$, since the input state is regular at $r=0$.

\begin{figure}[htb]
\includegraphics[width=3.25 in]{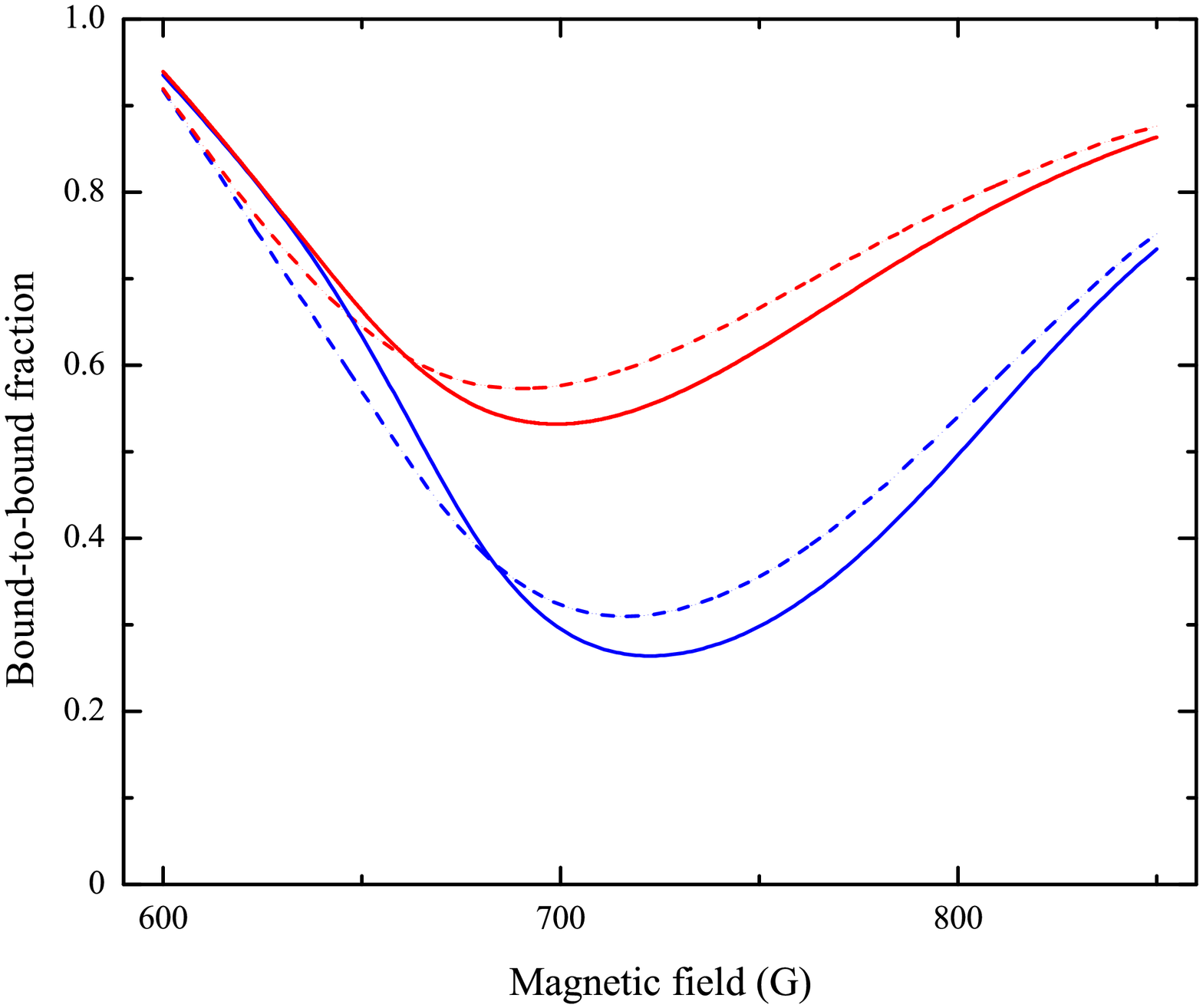}
\caption{Bound 12 dimer to bound 13 dimer transition fraction versus magnetic field for $\nu_\perp/\nu_z=1/25$. Upper curves (red online): Trap depth $U_0=280\,\mu$K, $\nu_z=82.5$ kHz; Lower curves (blue online) $U_0=21\,\mu$K and $\nu_z=24.5$ kHz.  Solid lines denote the square of Frank-Condon overlap integral, including the first 50 even axial states.  Dashed lines denote the corresponding results using only the ground axial state, Eq.~\ref{eq:boundtoboundfraction}.\label{fig:BoundtoBoundFraction}  }
\end{figure}

 For $2\rightarrow 3$ transitions in a 12 mixture at 834 G,  the binding energies are small compared to the energy difference between symmetric axial states $2\hbar\omega_z$, which are coupled by the s-wave scattering interaction. In this case, the Green's functions, and hence the 1-2 and 1-3 bound states and the 1-3 scattering states, are well approximated by the ground axial state component $\propto \int dz\phi_0(z)\,G_{\epsilon_b}(z,\rho)$, yielding the normalized bound state
 \begin{equation}
 \psi_{\epsilon_b}(z,\rho)=\phi_0(z)\,\frac{\kappa}{\sqrt{\pi}}\,K_0(\kappa\rho),
 \label{eq:bound}
 \end{equation}
where $\phi_0(z)$ is the axial ground vibrational state for the relative motion, $E_b=\epsilon_b\,\hbar\omega_z$ is the pair binding energy for the given scattering length, and $\hbar^2\kappa^2/m=E_b$, i.e., $\kappa=\sqrt{\epsilon_b}/l_z$.  The 1-2 bound state to 1-3 bound state transition strength is then determined by the overlap between of the modified Bessel functions, $K_0(\kappa_{12}\rho)$ and $K_0(\kappa_{13}\rho)$, where $\kappa_{12(13)}$ is determined by $\epsilon_b^{12(13)}$. The 1-2 bound to 1-3 bound contribution to the lineshape is then given by
\begin{equation}
I_{bb}(\nu)=\epsilon_{bb}(q)\,\delta[\nu-(\epsilon_b^{12}-\epsilon_b^{13})\nu_z],
\end{equation}
where $\nu_z$ is the axial harmonic oscillator frequency in Hz and $\nu$ is the rf frequency in Hz, relative to the bare 2-3 hyperfine transition frequency of $\simeq 83$ MHz. For the axial ground state, the frequency integrated bound to bound transition strength can be written compactly in terms of $q\equiv \ln(\epsilon_b^{13}/\epsilon_b^{12})$,
\begin{equation}
\int d\nu\,I_{bb}(\nu)\equiv\epsilon_{bb}(q)= \frac{q^2}{4 \sinh^2(q/2)}.
\label{eq:boundtoboundfraction}
\end{equation}
We plot $\epsilon_{bb}$ as a function of magnetic field in Fig.~\ref{fig:BoundtoBoundFraction} for the general case, valid for both weak and tight binding of the $1-2$ or $1-3$ dimers, including the contribution of the first 50 even axial states.
 For comparison, we show the weak binding approximation given by Eq.~\ref{eq:boundtoboundfraction}, which includes only the ground axial vibrational state. For small binding energies, increasing the trap depth significantly increases the pairing energy, increasing the overlap integral for bound-to-bound transitions.

 From Fig.~\ref{fig:BoundtoBoundFraction}, we see that for a trap depth $U_0=21\,\mu$K at 720 G, the $2\rightarrow 3$ transition in a 12 mixture is dominated by bound-to-free transitions. In contrast, at 834 G, the corresponding bound-to-bound transition is dominant and increases with increasing trap depth.

In the same approximation, the two dimensional box-normalized (to area A) 1-3 scattering state takes the form $\psi_{\epsilon_\perp}(z,\rho)=\phi_0(z)\,\psi_{\epsilon_\perp}(\rho)$, with $E_s=\epsilon_\perp\hbar\omega_z=\hbar^2\,k_\perp^2/m$. Assuming that the incident transverse state is the azimuthally-symmetric part ($l=0$) of a box normalized plane-wave state $\exp(i\mathbf{k}_\perp\cdot\mathbf{x}_\perp)/\sqrt{A}$, i.e., $J_0(k_\perp\rho)/\sqrt{A}$, Eq.~\ref{eq:scattering} gives
\begin{equation}
\psi_{\epsilon_\perp}(\rho)=\frac{1}{\sqrt{A}}\left\{J_0(k_\perp\rho)-\frac{\pi i}{\ln(\epsilon_b/\epsilon_\perp)+\pi i}\,H_0(k_\perp\rho)\right\}.
\label{eq:scatteringstate}
\end{equation}
We determine the overlap integral of the 1-2 bound state with the 1-3 scattering state and integrate the transition rate using  density of transverse states, $A/(2\pi)2\pi k_\perp dk_\perp$, to obtain the 1-2 bound to 1-3 scattering state  contribution to the lineshape, which takes the form of a threshold function,
\begin{equation}
I_{bf}(\nu)=\frac{\epsilon_b^{12}\nu_z}{\nu^2}\,
\frac{q^2\,\theta(\nu-\epsilon_b^{12}\nu_z)}{\left[q-\ln\left(\frac{\nu}{\epsilon_b^{12}\nu_z}-1\right)\right]^2+\pi^2}.
\label{eq:boundtofree}
\end{equation}
Note that $\int d\nu\,I_{bf}(\nu)=1-\epsilon_{bb}(q)$, as it should. We find that Eq.~\ref{eq:boundtofree} well fits the radio frequency spectra obtained at 720 G, where the dimer binding energy is larger than the transverse Fermi energy.

\section{Polarons in a two-dimensional Fermi gas}
\label{sec:polaron}
For radiofrequency $2\rightarrow 3$ spectra obtained near 834 G in a 12 mixture, we find that the difference between the calculated dimer binding energies significantly underestimates the observed frequency shifts, as shown in Figs. 2, 3, and 4 of the main paper. We consider the possibility that the spectra may arise from transitions between polaronic states, as the polaron energy is significantly more negative than the corresponding dimer energy and is therefore energetically preferred~\cite{Pethick2DPolarons}.

To estimate the polaron energies, we consider either an isolated impurity atom in state 2 or in state 3, immersed in a bath of atoms in state 1. We employ the method described for polarons in three dimensions in the supplementary material of Schirotzek at al.~\cite{MartinPolaron}, which is based on the zero momentum polaron wavefunction proposed by Chevy~\cite{ChevyPolaron}, which for a polaron in state $i=2,3$ takes the form

\begin{equation}
|E_i\rangle=\varphi_{0i}|0\rangle_i\,|FS\rangle_1+\hspace*{-0.1in}\sum_{q<k_F<k}\hspace*{-0.1in} \varphi_{\mathbf{k}\mathbf{q}}|\mathbf{q-k}\rangle_i\, c^\dagger_{\mathbf{k}1}c_{\mathbf{q}1}\,|FS\rangle_1.
\label{eq:polaronstate}
\end{equation}
Here, the first term describes an impurity $i$ of zero momentum in a Fermi sea of atoms in state $1$ for which the net momentum is zero. Collisions between the impurity and the background atoms couple the zero momentum impurity state to that with momentum $\mathbf{q-k}$, producing a particle-hole pair from the Fermi sea of atoms in state $1$ with net momentum  $\mathbf{k-q}$, conserving the total zero momentum.

For the 2D calculations, we replace the box normalization volume in the supplementary material of Schirotzek at al.~\cite{MartinPolaron} by the corresponding area $A$, so that the polaron energy in 2D takes the form
\begin{equation}
E_{pi}=\frac{1}{A}\sum_{q<k_F}f(E_{pi},q),
\label{eq:polaronenergy1}
\end{equation}
where
\begin{equation}
f^{-1}(E_{pi},q)=\frac{1}{g_0}+\frac{1}{A}\sum_{k>k_F}\frac{1}{\epsilon_\mathbf{k}-\epsilon_\mathbf{q}+\epsilon_\mathbf{q-k}-E_{pi}}.
\end{equation}
Here, $\epsilon_k=\hbar^2k^2/(2m)$ and $k_F$ is the local Fermi wavevector with $E_F=\hbar^2k_F^2/(2m)$ the corresponding local transverse Fermi energy.

 Following Z\"{o}llner et al., Ref.~\cite{Pethick2DPolarons}, we assume that the effective bare interaction $U$ arises from a short range 2D potential, so that the matrix elements $U_{\mathbf{k}\mathbf{k}'}=g_0/A$ are momentum independent. $g_0$  can be rewritten using the physical two-body T-matrix element in 2D, $1/g_0=1/T_{2B}(\mathbf{k}_0)-(1/A)\sum_\mathbf{k}1/(2\epsilon_k-2\epsilon_{k_0})$. This method is similar to that employed previously~\cite{Randeria2D,Pethick2DPolarons,Parish2DPolarons}. Here, we choose the T-matrix element $T_{\mathbf{k}'\mathbf{k_0}}=T_{2B}(\mathbf{k}_0)/A$ so that the scattering rate calculated using the generalized Golden rule reproduces the scattering rate obtained from the 2D flux corresponding to Eq.~\ref{eq:scatteringstate}. Then $T_{2B}(\mathbf{k}_0)=\hbar^2\,f(k_0)/m$, with $f(k_0)=4\pi/[\pi i+ln (E_b/\epsilon_\perp)]$. Here, $\epsilon_\perp=2\epsilon_{k_0}$ and $E_b=\epsilon_b\hbar\omega_z$ is the dimer binding energy calculated from Eq.~\ref{eq:bindingenergy}. Both the $\pi i$  term and the $k_0$ dependence in $T_{2B}(\mathbf{k}_0)$ are canceled by corresponding terms in the sum $(1/A)\sum_\mathbf{k}1/(2\epsilon_k-2\epsilon_{k_0})$, so that $f^{-1}(E_{pi},q)$ is independent of $k_0$ as it should be.  For attractive polarons with energy $E_{pi}<0$,  we then obtain the simple integral equation $\epsilon(L_i)=\Sigma(L_i,\epsilon_i)$, where
\begin{equation}
\Sigma\equiv\int_0^1\frac{-2\,du}{-L+ln[\sqrt{(1-\frac{\epsilon}{2})^2-u}+(1-\frac{\epsilon}{2}-\frac{u}{2})]}.
\label{eq:polaronenergy}
\end{equation}
Here,  $\epsilon(L_i)\equiv\epsilon_i= E_{pi}/E_F$ and $L_i=ln(E_b^{1i}/E_F)$ for an impurity in state $i$.
Eq.~\ref{eq:polaronenergy} yields the polaron energies $E_{pi}=\epsilon_i\,E_F$ for the initial and final states $i=2,3$, using the binding energies $E_b^{1i}=\epsilon_b^{1i}\hbar\omega_z$ determined from Eq.~\ref{eq:bindingenergy}.

\begin{figure}[htb]
\includegraphics[width=3.25in]{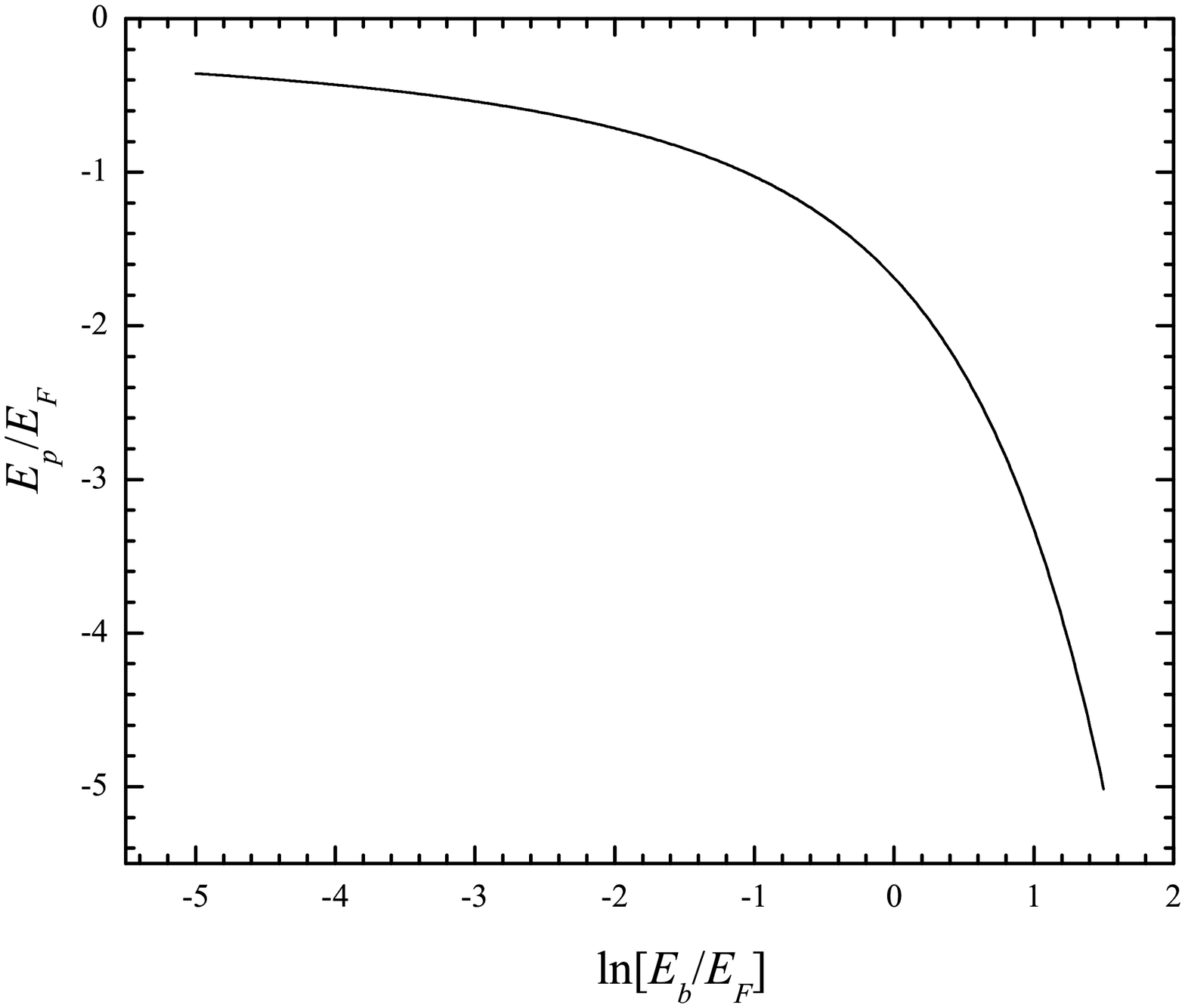}
\caption{Polaron energy $E_p/E_F$ versus $\ln[E_b/E_F]$, where $E_b$ is the dimer binding energy and $E_F$ is the local Fermi energy.\label{fig:polaronenergy}}
\end{figure}

Fig.~\ref{fig:polaronenergy} shows the attractive polaron energies obtained from Eq.~\ref{eq:polaronenergy}, which agree with those obtained in Ref.~\cite{Pethick2DPolarons}. We see that the polaron energy $E_p$ is a large fraction of the local Fermi energy, which can be much larger than the corresponding dimer binding energy $E_b$.

\begin{figure}[t]
\includegraphics[width=3.0 in]{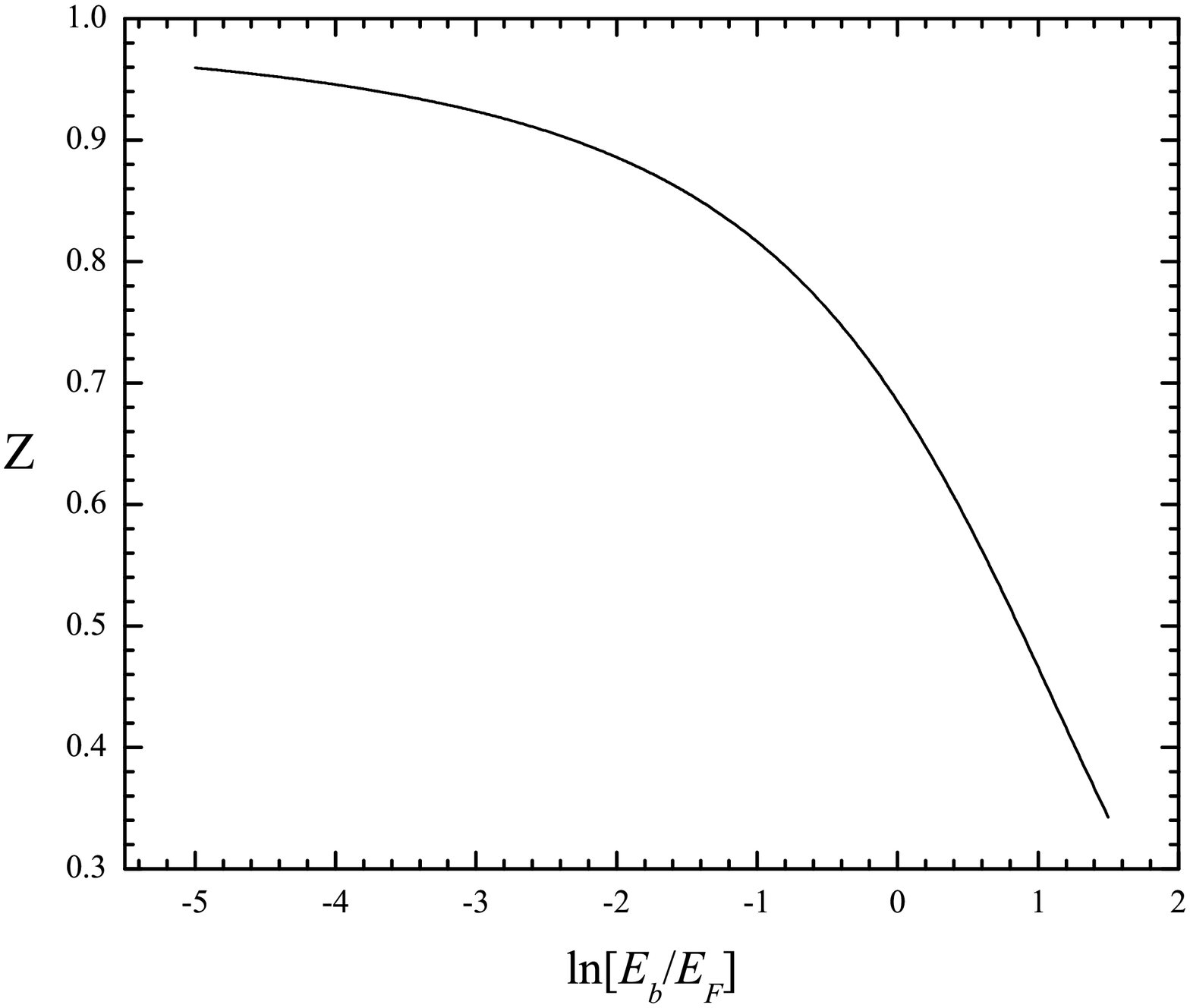}
\caption{Polaron quasiparticle weight $Z$ versus $\ln[E_b/E_F]$, where $E_b$ is the dimer binding energy and $E_F$ is the local Fermi energy.\label{fig:polaronweight}}
\end{figure}

For  radiofrequency transitions between impurity states $2\rightarrow 3$ in a bath of atoms in state 1, the momentum of the impurity does not change. We therefore assume the coherent part of the spectrum is given by
\begin{equation}
I(\hbar\omega)=Z_2\,Z_3\,\delta[\hbar\omega-E_{p3}+E_{p2}],
\label{eq:polaronspectrum}
\end{equation}
where $|\varphi_{03}^*\varphi_{02}|^2=Z_2\,Z_3$ is the square of the overlap integral between the part of the initial and final polaron states that yields the coherent part of the spectrum.  We determine $Z_2=|\varphi_{02}|^2$ using $Z_2^{-1}=1-\partial\Sigma(L_2,\epsilon)/\partial\epsilon$, with $\epsilon\rightarrow E_{p2}/E_F$, as described in Ref.~\cite{MartinPolaron} for the 3D problem,  and similarly  for $Z_3=|\varphi_{03}|^2$.
Fig.~\ref{fig:polaronweight} shows the quasiparticle weight $Z$ obtained from Eq.~\ref{eq:polaronenergy}.
For the dimer binding energies in a 12 mixture near 834 G, Table 1 of the main paper, we find that $Z_2\simeq 0.85$ and $Z_3\simeq 0.94$ for the shallowest trap depth, both close to unity.  Hence, we expect that the overlap between the initial and final polaron states is strong and that transitions between polaron states should make an important contribution to the spectrum.

In the limit that the dimer binding energy  is small compared to the local Fermi energy, i.e., $E_b<<E_F$, one verifies from Eq.~\ref{eq:polaronenergy} that the corresponding polaron energy yields the limiting form~\cite{Pethick2DPolarons}
\begin{equation}
E_p\simeq \frac{-2\,E_F}{\ln(2\,E_F/E_b)},
\label{eq:MeanField2D}
\end{equation}
which can be interpreted as an effective mean field shift, since $E_F$ is proportional to the 2D density $n_\perp(\rho)$. At 842 G, where the dimer binding energy is reasonably small compared to the transverse Fermi energy in our experiments, this formula overestimates the magnitude of the polaron energy difference $E_{p3}-E_{p2}$ by about 10\% for the most shallow trap.  At 811 G, it overestimates the energy difference by about 40\%. As $E_b$ is not small compared to $E_F$ for most of the data, it is not surprising that $E_F$ cannot be adjusted in Eq.~\ref{eq:MeanField2D} to give the measured frequency differences.
